\documentstyle[12pt, fleqn]{article}
\def\ref{\par\noindent\hangindent=6mm\hangafter=1}
\def\x{{\bf x}}

\begin{document}

\baselineskip 8mm
\

\vspace{10mm}

\begin{center}

{\bf Abundance and Clustering of C IV Absorption Systems

 in the SCDM, LCDM and CHDM Models}

\bigskip
\bigskip
\bigskip
\bigskip

Hongguang Bi$^*$ and Li-Zhi Fang

\bigskip

Department of Physics, University of Arizona

Tucson, AZ 85721

\end{center}

\vspace{4cm}

{\it All correspondence to: L.Z. Fang}

\newpage

\begin{center}

{\bf Abstract}

\end{center}

\bigskip

We have developed a method for calculating the two-point correlation
function of nonlinearly evolved mass and collapsed halos in the
Press-Schechter formalism. The nonlinear gravitational interaction is
treated as the sum of various individual spherical top-hat clustering.
Because no collapsed halo of mass $M$ can exist in initial regions
(or top-hat spheres) of mass less than $M$, the bias that massive halos
have stronger correlation than the background mass can be naturally
introduced.

We apply this method to derive constraints on popular dark-matter
models from the spatial number density and the correlation function of C IV
absorption systems in QSO spectra. Considering C IV systems should be
hosted by collapsed halos, one can obtain an upper limit to the threshold
mass of the collapsed halos by requiring their number density to be
larger than that of observed C IV systems.
On the other hand, in order to explain the observed clustering of C IV
systems,
a lower limit to the threshold mass will be set for the hosting
halos. We found that the standard cold dark matter (SCDM) model and
the low-density flat universe with a cosmological constant $\Lambda_0$
(LCDM)
are consistent with the abundance and clustering of C IV systems. However,
the two cold-plus-hot dark matter models (CHDMs) with the cosmological
parameters ($\Omega_c+\Omega_b)/\Omega_h=0.7/0.3$ and $0.8/0.2$,
respectively, have difficulty passing
the two tests simultaneously. In these models, in order
to have enough collapsed halos to host C IV systems, the threshold mass of
the halos can not be greater than $10^{11}\ M_{\odot}$.  But in order to
agree with the two-point correlation function on the scales of $\Delta v
\sim 300 - 1,000$ km/s, the threshold mass should be larger than
$10^{12}\ M_{\odot}$.

\bigskip
\bigskip
\bigskip

\noindent{\bf Subject headings:} cosmology: theory --- dark matter
--- quasars:
  absorption lines

\newpage
\

\vspace{4mm}

\noindent{\bf 1. Introduction}

\bigskip

It has been recognized that the abundance, i.e. the spatial number density,
of moderate and high redshift objects can put promising constraints on
models of structure formation in the universe. For instance, an N-body
simulation has shown that the $``$standard" cold-plus-hot dark matter model
(CHDM) of $\Omega_c/\Omega_h/\Omega_b= 0.6/0.3/0.1$, here $c, b$ and $h$
denoting cold, baryonic and hot respectively, lacks the perturbations
necessary to form clusters before $z \ge 0.5$, although it can produce
a proper number of clusters at $z=0$ (Jing \& Fang 1994). In contrast, the
low density, flat cold-dark-matter model (LCDM) can produce enough clusters
at both $z\ge 0.5$ and $z=0$. Hence, the abundance of $z\ge 0.5$
clusters is a useful quantity for discriminating between the CHDM model and
the LCDM model. Another example is the damped Ly$\alpha$ systems which
are rich in QSO absorption spectra at $z\geq 2$. Both the
Press-Schechter formalism and N-body simulations indicated that the
$``$standard" CHDM model has difficulty explaining these systems (Mo \&
Miralda-Escud\'e 1994; Ma \& Bertschinger 1994; Klypin et al. 1995). A
simulation of the Ly$\alpha$ forest has shown that the predicted number of
forest lines is less than what is observed by at least one order of magnitude
even if extreme parameters are used in this model (Bi, Ge \& Fang 1995).

Among the various samples of high-redshift objects, those selected from
absorption systems in QSO spectra are relatively uniform and numerous and
can be used to provide stringent constraints on different theoretical
models. However, a common problem of 1-D samples is the lack of
information on the size
and geometry the objects. Calculating 3-D abundances from 1-D
data requires assumptions about the objects'
shape. Obviously, this leads to an
uncertainty in the results. This difficulty is partly overcome in this study.
We propose to test models not only by abundance, but also by clustering
property of a high-redshift sample. Like abundance, clustering of collapsed
halos depends also on their mass and/or size, so adding a clustering test
will reduce the unknown parameters. In some cases, conclusions
could even be completely free from these parameters.

The present analytical study is based on the Press-Schechter formalism
(Press \& Schechter 1974, hereafter PS) that has been found to be successful
in describing the mean number density of collapsed virial halos with
different mass threshold at high redshifts. The spatial range in our study
covers from redshift 2 to 4, and the correlation scales are from 300
to tens thousand km s$^{-1}$, which are otherwise difficult to handle by
current gas simulations.

For our goal, we need a method to calculate the correlation function of
collapsed halos in the PS formalism.
The existing PS theory does not yet tell us about the correlation function.
In the first part of this paper, we will discuss how to extend the PS
formalism to obtain the spatial correlation function of nonlinearly evolved
mass and collapsed halos. A similar problem has been studied recently
by Mo \& White (1995). The idea is that the gravitational clustering from
given initial density fluctuations can be approximately treated as many
individual top-hat-evolved spherical regions. The mass correlation can then
be deduced from the mass distribution within such regions. The
bias of halo autocorrelations with respect to their masses is derived by
assuming that no halo of mass $M$ can exist in uncollapsed regions of mass
less than $M$. This approximation is found to be in good agreement with
the linear approximation on large scales, and to be consistent with the
empirical formalism on scales where the non-linear effects are significant
(e.g. Hamilton et al. 1991).

In the second part, we apply the developed method to C IV absorption
systems in QSO spectra. We compare the abundance and clustering of
collapsed halos in theoretical models with those given by real C IV
observations. Among high-redshift absorption samples, only metal absorption
systems and Ly$\alpha$ forest lines can provide the
statistics of both linear number densities and spatial correlation
functions (Sargent, Boksenberg \& Steidel 1988). Ly$\alpha$ absorption
lines are most likely from clouds with large size and low column density
which are neither virialized nor completely confined (Bechtold et al. 1994;
Dinshaw et al. 1994). Hence, we should not simply identify their hosts as
collapsed halos. In contrast, metal systems are generally
believed to be from huge halos surrounding galaxies (Wolfe 1993).
Because stars were certainly formed in the systems, they must have
undergone non-linear collapse. Therefore, metal absorption lines are
probably the only available 1-D sample for our purpose.

Observationally, metal absorption systems contain Mg II selected
systems, C IV selected systems, and Lyman limit systems. However, the
categories of absorbers identified by different systems are not orthogonal
with each other (Wolfe 1993). The Lyman limit systems are almost
indistinguishable from the Mg II systems, and most Mg II systems exhibit C IV,
too. Most metal systems were obtained from ground observations. The C IV
systems are usually found in the redshift interval $1.2 \leq z \leq 4.1$,
and the MgII systems in $0.2  \leq z \leq 1.9$. Because we are interested
mainly in the physics at high redshift ($z >2$), only C IV absorption
systems will be investigated.

It should be pointed out that, like other studies of high-redshift objects
based on the PS formalism, our goal is not to model the details of the
C IV systems, instead it is to use the number of C IV systems as a
{\em lower limit} to the number of corresponding collapsed halos.
It is believed that the
presence of C IV depends on the chemical abundance of heavy elements and on
the ionization state of baryonic gas (Bergeron et al. 1994). Observations
have shown an evolution in these chemical properties (Steidel 1990). It
indicates that C IV systems should be located in areas in which chemical
abundance evolution has already taken place. Therefore, C IV systems should
be harbored in collapsed PS halos. Thus, a reasonable
constraint on models of structure formation is
that the predicted number of collapsed PS halos should be greater than
that of observed C IV systems.

In \S 2, we describe the PS method for calculating the spatial number density
of collapsed halos, then we extend this idea to general uncollapsed regions
and derive an approximate expression for the correlation functions of mass
and halos. In \S 3, the four SCDM, LCDM and CHDM models of structure
formation are tested based on their predictions of the abundance and
clustering of collapsed halos. We discuss how observations can be compared
with the predictions. Finally, \S 4 gives discussion and conclusions.

\bigskip

\noindent{\bf 2. Method}

\bigskip

\noindent{\it 2.1 Number density of collapsed halos}

\bigskip

The spatial number density of collapsed halos can be calculated from
the standard PS theory.
We define $\delta (\x, z)$ to be the 3-D density fluctuation field of
dark matter extrapolated to redshift $z$ assuming linear evolution.
A density field $\delta_R(\x)$, representing the smoothed fluctuation on
scale $R$, can be derived from $\delta (\x)$ by
\begin{equation}
\delta_R (\x ) = \frac{1}{V_R} \int \delta (\x_1) W(R; \x_1 -\x ) d\x_1,
\end{equation}
where the function $W(R; \x_1 -\x)$ is the top-hat window for the comoving
volume $V_R = 4\pi R^3/3$. In the $\Omega =1$ Einstein-de Sitter universe,
the variance of $\delta_R$ evolves as $\sigma^2_R \propto (1+z)^{-2}$.
The total mass within $V_R$ is $M = V_R \rho_0$, where $\rho_0$ is the
present
cosmological density when the scaling factor of the universe is set to be
unity at $z=0$.

For a Gaussian field,
the fraction of the total mass $\rho_0 \Delta \x$ having fluctuations
larger than a given $\delta _c$
in an arbitrary spatial domain $\Delta \x $ is
\begin{equation}
F_R = \int ^\infty_{\delta_c} \frac{1}{\sqrt{2\pi} \sigma_R}
      \exp{\left(-\frac{\delta_R ^2}{2\sigma_R ^2}\right)} d\delta_R.
\end{equation}
Therefore, if we take $\delta_c = 1.686$, the critical value for
collapse in the top-hat evolution, $F_R \cdot \rho_0 \Delta \x$
should be identified as the sum of masses of all collapsed halos, each
of which is greater than $M = V_R \rho_0$. The differential
$-\frac{\partial}{\partial M} (F_R \rho_0 \Delta \x ) dM$
gives the total mass of collapsed halos in the range $M$
to $M + dM$. Hence, if we define $n_c(M) dM$ to be the spatial
number density of halos between $M$ and $M+dM$, we have
\begin{equation}
-\frac{\partial}{\partial M} (F_R \rho_0 \Delta \x ) dM
=n_c(M) dM \cdot M\Delta \x ,
\end{equation}
where we use the subscript $c$ in $n_c$ to emphasize that it is for
collapsed halos.

Because the cloud-in-cloud problem has not been appropriately
considered in Eq. (3),
the original PS theory takes an ad hoc assumption
that the above defined
$n_c$ should be multiplied by a factor of 2 (see discussions in
Bond et al. 1991).
The normalization $\int_0^{\infty} n_c(M) M dM = \rho_0$ can thus
be fulfilled when $\sigma_R (R\to 0) = \infty$. One has finally
\begin{equation}
n_c (M) = -\frac{\rho_0}{M} \frac{\partial}{\partial M}
           {\rm erfc}\left(\frac{\delta _R}{\sqrt{2}\sigma _R}\right),
\end{equation}
where erfc(x) is the complementary error function. The cumulative number
density, $N_c(M)$, of all halos greater than $M$ should be
\begin{equation}
N_c (M ) = \int ^{\infty} _M n_c (M_1 ) dM_1.
\end{equation}
The abundance of halos calculated from Eq. (5) has been
verified by a number of N-body simulations (e.g. Efstathiou et al. 1988;
Lacey \& Cole 1994; Jing \& Fang 1994).

\bigskip

\noindent{\it 2.2 \ $\delta_c$ in the flat universe with a cosmological
constant}

\bigskip

The critical value of collapse, $\delta_c = 1.686$, is derived
in the Einstein-de Sitter model. Now, we calculate $\delta_c$
in the flat $\Lambda \neq 0$ universe. In this case, the evolution of a
spherical volume with mass $M$ should be described by a Newtonian equation
in the proper coordinate (Peebles 1984) :
\begin{equation}
\frac{d^2 r}{dt^2} =   - \frac{G M}{r^2} +\frac{1}{3}\Lambda r ,
\end{equation}
where $\Lambda = (1-\Omega)3 H_0 ^2 $. Eq.(6)
can be integrated to give
\begin{equation}
\left(\frac{dr}{dt} \right) ^2 = 2GM(\frac{1}{r} -\frac{1}{r_m}) +
                    (1-\Omega )H_0 ^2 (r^2 -r_m ^2),
\end{equation}
where $r_m$ denotes the radius when the sphere reaches its maximum size.
 From Eq. (7), one has
\begin{eqnarray*}
t & = & \int ^r _0 [ 2GM(\frac{1}{r} -\frac{1}{r_m}) +
                    (1-\Omega )H_0 ^2 (r^2 -r_m ^2) ]^{-1/2} dr,
\end{eqnarray*}
\begin{equation}
\ \ \ \ = \ \frac{1}{\sqrt{1-\Omega} H_0} \int ^{r/r_m} _0
    y^{\frac{1}{2}} (1-y)^{-\frac{1}{2}} (c_0 -y^2 -y)^{-1/2} dy,
\end{equation}
where $c_0 \equiv {2GM}/{(1-\Omega )H_0^2 r_m ^3}$.

If we define $t_m$ to be the cosmic time at which the sphere is at its
maximum radius $r_m$, the collapse time $t_c$ of this sphere
is twice $t_m$.
Therefore, taking $t_m = t_c/2$ in Eq. (8), we find
\begin{equation}
t_c(c_0) = \frac{2}{\sqrt{1-\Omega} H_0} \int ^{1} _0
    y^{\frac{1}{2}} (1-y)^{-\frac{1}{2}} (c_0 -y^2 -y)^{-1/2} dy.
\end{equation}
The redshift $z_c$ that is a function of $c_0$ can be found according to
the $z - t$ relationship in the flat universe:
\begin{equation}
H_0 t = \frac{2}{3(1-\Omega )^{1/2}} \sinh ^{-1} [\sqrt{\frac{1-\Omega}
{\Omega}}(1+z)^{-3/2} ].
\end{equation}

At the time of recombination $t_d$, the radius of the sphere $r_d$ is
much less than $r_m$, so the integration in Eq. (8)
can be approximated as
\begin{equation}
t_d = \frac{1}{\sqrt{1-\Omega} H_0} \left[\frac{2}{3} (\frac{r_d}{r_m})^{3/2}
    + \frac{1+c_0}{2c_0} \frac{2}{5} (\frac{r_d}{r_m}) ^{5/2} \right ].
\end{equation}
Again using the $z- t$ relationship, one can write
the density perturbation in the sphere at $t_d$,
$\delta _{d} = M/\frac{4\pi}{3}r^3_{d} (1+z_d)^3 \rho_0 -1$, as
\begin{equation}
\delta_{d}(c_0) =
  \frac{3c_0(1-\Omega )H_0^2 r_m ^3}{8G\pi r^3 _{d} (1+z_d)^3 \rho _0} -1
  \simeq \frac{1+c_0}{c_0} \frac{3}{5}
  \left[\frac{(1-\Omega)c_0}{\Omega}\right]^{1/3} \frac{1}{1+z_d}.
\end{equation}
Since $z_c$ depends only on $c_0$, we can write the above $c_0$ as
a function of $z_d$, i.e. $c_0 = c_0 (z_d)$.

Using this initial density fluctuation, one
obtains the subsequent linear evolution at the epoch of collapse $t_c$ :
\begin{equation}
\delta_c(z_c)  =  D(z_c)(1+z_d)\delta_d  =
\end{equation}
$$   [\Omega (1+z_c)^3 +1-\Omega ]^{1/2}
     \frac{3}{5} \frac{1+c_0(z_c)}{c_0(z_c)}
    \left [\frac{(1-\Omega)c_0(z_c)}{\Omega} \right ]^{1/3}
 \int ^{\infty}_{z_c}\frac{(1+z) dz}{[\Omega (1+z)^3 +1-\Omega ]^{3/2}} $$
where $D(z)$ is the linear growing factor in the LCDM model.

Fig. 1 shows $\delta_c(z_c)$ vs. $z_c$ for $\Omega = 0.9$, 0.7, 0.5, 0.3,
0.1. All of the $\delta_c(z_c)$ curves are approaching the traditional
value 1.686; there is no noticeable difference among the thresholds at
redshifts $z\ge 1$.  This indicates that the trajectory of the
top-hat collapse in the flat $\Lambda$ universe is very well
described by simply the Einstein-de Sitter universe at high redshift,
so we will take $\delta_c = 1.686$ in the following calculations.

\bigskip

\noindent{\it 2.3  Number density of uncollapsed regions}

\bigskip

As pointed out by Mo \& White (1995), the nonlinear effect of gravitational
clustering of dark matter can be treated as the sum of various individual
top-hat spheres including both collapsed halos and uncollapsed regions.
An uncollapsed region (sometimes called {\em PS spherical region} or {\em
uncollapsed sphere}; {\em halo} is usually a terminology for collapsed
objects) corresponds to linear fluctuations not yet reaching the
threshold 1.686 at the redshift considered. Therefore, we can apply the same
statistical method as that used by Press and Schechter
for collapsed halos to uncollapsed regions. Following \S 2.1,
we first calculate the number density of uncollapsed regions.

The evolution of a spherical region of radius $r = r_d$ at
recombination $z=z_d$
is described by Eq. (6). In the Einstein-de Sitter case,
the solution of Eq. (6) can be expressed analytically as
\begin{equation}
r = \frac{3}{10} \frac{r_0}{\delta _0} (1 - \cos{\theta}),
\mbox{\ \ \ \ \ \ }
\frac{1}{1+z} = \frac{3\times 6^{2/3}}{20\delta_0}
               (\theta - \sin{\theta} )^{2/3}
\end{equation}
for $\delta_0 > 0$, and
\begin{equation}
r = \frac{3}{10} \frac{r_0}{\delta _0} (1 - {\rm cosh }{\theta}),
\mbox{\ \ \ \ \ \ }
\frac{1}{1+z}  = - \frac{3\times 6^{2/3}}{20\delta_0}
               ({\rm sinh }{\theta} - \theta )^{2/3}
\end{equation}
for $\delta_0 < 0$, where $r_0 \equiv (1+z_d) r_d$ and $\delta_0 \equiv
(1+z_d)\delta _d$ are, respectively, the linear extrapolation of the radius
and density contrast to $z=0$. Note that the mass of this sphere is then
given by $M=\frac{4\pi}{3} (1+z_d)^3 r_d ^3 \rho_0(1+\delta_d)$.

Let's consider an arbitrary spherical volume of radius $r$ at redshift $z$.
Matter in this volume can come from various initial spheres of different
$r_0$ and $\delta_0$. For a given $r$ and $z$, one can find the relationship
between $r_0$ and $\delta_0$ from Eqs. (14) and (15). This is
$\frac{\delta_0}{\delta_c (1+z)} = f(\frac{r(1+z)}{r_0})$. Function $f(x)$
is plotted in Fig. 2. The meaning of the figure is straightforward. When
$\delta_0=0$, one has $r=r_0/(1+z)$, i.e. the evolution of this region is
just comoving. An initial sphere with perturbation $\delta _0>0$
will evolve into the radius $r < r_0/(1+z)$ at redshift
$z$. For $\delta _0 < 0$, it will evolve into $r > r_0/(1+z)$. All
initial spheres with $\delta_0 > \delta_c(1+z)$ result in
$r =0$, i.e.  they are collapsed halos.

One can consider the initial density field as a system consisting of many
spheres, each of which has the same
radius $r_0$ but various density contrasts $\delta_0$. Similar to
Eq. (2), the mass fraction of the spheres greater than
$\delta_0$ is given by
\begin{equation}
\int ^\infty _{\delta _0} \frac{1}{\sqrt{2\pi} \sigma _{r_0}}
      \exp{\left(-\frac{\delta ^2}{2\sigma _{r_0} ^2}\right)} d\delta
\end{equation}
where $\sigma_{r_{0}}^2$ is the variance of density perturbation on
scale $r_0$. Because a spherical region of $\delta_0$ and $r_0$ will
evolve into $r$ at $z$, initial spherical regions of $\delta \ge \delta _0$
and $r_0$ will evolve into radii less than $r$ at $z$. Therefore, the
comoving number density of the uncollapsed PS spheres with mass in
$M_0 \to M_0 + dM_0$ (or radii $r_0\to r_0 +dr_0$) at recombination, which
are spheres with radius less than $r$ at $z$, is given by
\begin{equation}
n(M_0) dM_0 = -\frac{\rho_0}{M_0} \frac{1}{2} \frac{\partial}{\partial M_0}
   {\rm erfc}\left[\frac{\delta_0}{\sqrt{2}\sigma (r_0)}\right] dM_0,
\end{equation}
where we use the subscript $0$ in $M_0$ to emphasis that it is for the
uncollapsed regions, not only the collapsed halos discussed in \S 2.1.
It should be pointed out that Eq. (16) contains spheres of both
$\delta_0 < \delta_c(1+z)$ and $\delta_0 > \delta_c(1+z)$.
Therefore, $n(M_0)dM_0$, in fact, includes both uncollapsed regions and
collapsed halos. Since initial spherical regions entirely cover the density
field, $n(M_0)$ should satisfy the normalization condition
$\rho  = \int ^{\infty}_0 n(M_0 )M_0 dM_0 = \rho_0$ for all $r$ and $z$.
It is easy to verify this normalization from Eqs. (16) and (17).

\bigskip

\noindent{\it 2.4 Mass-correlation function}

\bigskip

Since $n(M_0) dM_0$ gives the number density of uncollapsed
spherical regions with mass $M_0 \to M_0 + dM_0$ and
radii $ \leq r$ at $z$, the fraction of regions with radius
$r \rightarrow r + dr$ can be obtained by
differentiating $n(M_0)$ with respect to $V=4\pi r^3/3$.
Defining $m(M_0, V) dM_0 dV$ to be the number density of the regions
with radius $r \rightarrow r + dr$ or $V \rightarrow V + dV$,
where $dV=4\pi r^2 dr$, we have
\begin{eqnarray*}
m(M_0, V) & = &  \frac{\partial}{\partial V} n(M_0)
\end{eqnarray*}
\begin{equation}
 \ \ \ \ \ \ \ \ \ \ \ \ \ \ \ \ = \
    -\frac{\rho_0}{M_0} \frac{1}{2} \frac{\partial ^2}
             {\partial M_0 \partial V}
    {\rm erfc}\left[\frac{\delta _0}{\sqrt{2}\sigma (r_0)}\right].
\end{equation}
Therefore, the total number of such spherical regions
in an arbitrary volume $dV_1$ should on average be
given by $m(M_0, V) dM_0 dV dV_1$.

The mass correlation function $\xi(r)$ can be defined as
the relative enhancement of mass density in the spherical shell
$r \rightarrow r + dr$ around $dV_1$. Only the spheres with radius
$r \rightarrow r + dr$ can contribute to this enhancement.
The mean enhancement of each $M_0$ sphere is approximately
described by its mass variance $M_0^2\sigma ^2(M_0, z)$, where
$\sigma^2$ is the variance of the linear density contrast extrapolated
to $z$. Therefore, the mass correlation function can be
expressed as
\begin{eqnarray*}
\xi(r, z) & = & \frac{\int_0 ^{\infty} m(M_0, V) dM_0 dV dV_1 \cdot M_0 ^2
          \sigma ^2 (M_0, z)}{\rho_0 dV \cdot \rho_0 dV_1}
\end{eqnarray*}
\begin{equation}
\ \ \ \ \ \ \ \  = \ \int_0^{\infty} dM_0 m(M_0, V) V_0 ^2 \sigma^2
(M_0, z).
\end{equation}
The factors $\rho_0 dV$ and $\rho_0 dV_1$ are the mean mass in the
spherical shell $r \rightarrow r + dr$ and the volume $dV_1$, respectively.
Obviously, in deriving Eq. (19), we assumed that there is
no correlation among the initial spheres, so the mass fluctuation in the
shell is simply given by the sum of the individual components.

We can check the approximation of Eq. (19) by comparing it with the linear
approximation and other empirical non-linear formulae.
First, because the mass correlation function is very well described
by the linear approximation on large scales, Eq. (19)
should be equal to the linear correlation function when $r$ is large.
The integrated mass correlation function $\overline{\xi}$ that is
defined by
\begin{equation}
\overline{\xi} (r)
= \frac{1}{V} \int^V _0 \xi (V_1) d V_1, \mbox{\ \ \ \ \ \ \ }
\xi (r)  = \frac{\partial}{\partial V} (V\overline{\xi} (r))
\end{equation}
is shown in Fig. 3 for the  SCDM, LCDM, CHDM1 and CHDM2
models respectively at $z=2.8$. The solid lines in Fig. 3
represent our integrated
correlation function from Eq.(19).  The dotted lines in Fig. 3 are the
linear approximation that matches Eq. (19) exactly on large scales.

The dashed line at the upper left of Fig. 3, i.e. in the SCDM model,
is the empirical correlation function of
Hamilton et al. (1991) fitted to early N-body simulation results in
the SCDM model. Our approximation has nonlinear clustering behavior
similar to
theirs except for higher correlations on small scales. Since there has been
discussion as to whether early N-body simulations underestimated the
correlation function on the smallest scales, a further judgment on
these two approximations should be done by simulations with higher
resolution.

\bigskip

\noindent{\it 2.5  Correlation functions of collapsed halos}

\bigskip

We now consider the correlation function of collapsed halos.
 From Eq. (5), the number of collapsed halos in an arbitrary volume
$V_0$ is $N_c(M) V_0$. However, this is not the number {\em within}
uncollapsed spheres which we discussed in the last two
sections, because it is impossible that
a collapsed halo can form from uncollapsed regions if the mass of the
initial sphere is less than the mass of the collapsed halo.

To avoid this difficulty, we require that inside an
uncollapsed sphere of mass $M_0$, the number density of collapsed halos
is zero if their mass $M$ is greater than $M_0$, or is proportional
to $N_c$ if their mass $M$ is less than $M_0$.
The number density of collapsed halos inside an PS sphere is
thus given by
\begin{equation}
{\cal{N}} _c (M, M_0) = \left\{ \begin{array}{ll}
                        A N_c(M) & \mbox{for $M \le M_0$,} \\
                        0            & \mbox{for $M > M_0$,}
                        \end{array}
                     \right.
\end{equation}
where the constant $A \ge 1$ is introduced to maintain the normalization
condition
\begin{equation}
\int ^{\infty} _0 {\cal{N}} _c V_0 n(M_0 ) dM_0 = N_c.
\end{equation}
Therefore, we have $A=2/ {\rm erfc} [\delta _0(r_c)/\sigma (r_c)]$
and $r_c = (3M/ 4\pi \rho_0)^{1/3}$.

Obviously, Eq. (21) is a simplified description of geometric bias
(Kaiser 1984). Since the whole density field can be covered by spherical
regions, every halo should be contained in a proper sphere.
Massive collapsed halos can form
only in massive uncollapsed spheres. Therefore, the probability of
finding a collapsed halo in a massive uncollapsed sphere should be higher
than in an average arbitrary volume by the factor $A \ge 1$. This effect will
enhance the spatial correlation of massive halos.

As in \S 2.4, we consider a typical spherical shell
$dV=4\pi r^2dr$. The total number of uncollapsed regions with radii
$r \rightarrow r+dr$
and masses $M_0 \rightarrow M_0 +dM_0$ in the volume $dV_1$
is $m(M_0, V) dM_0 dV dV_1$. The total number of
collapsed halos greater than $M$
in each $V_0$ sphere is ${\cal{N}}_c(M) V_0$, and the variance of the
number is
$({\cal{N}} _c V_0)^2 \sigma ^2 (r_0)$. Therefore, the correlation
function of collapsed halos with mass larger than $M$ can be
expressed as
\begin{eqnarray*}
\xi (r; >M) & = & \frac{\int _M ^{\infty} dM_0 m(M_0, V)  dV dV_1
          \cdot ({\cal{N}} _c V_0)^2
          \sigma ^2 (M_0, z)}{dV N_c \cdot dV_1 N_c}
\end{eqnarray*}
\begin{equation}
\ \ \ \ \ \ \ \ \ \ \ \ = \
    A^2 \int _M ^{\infty} dM_0 m(M_0, V) V_0 ^2 \sigma ^2 (M_0, z)
\end{equation}
where we implicitly assume that in uncollapsed regions, the collapsed
halos have the same linear variance as the mass.

\bigskip

\noindent{\bf 3. Application to C IV systems}

\bigskip

We will concentrate on four models of structure formation:
SCDM, LCDM, CHDM1 and CHDM2. The parameters of these models are listed in
Table 1, including the Hubble constant $h$ in the unit of 100
km s$^{-1}$ Mpc$^{-1}$,
the cosmological density parameters $\Omega$, the cosmological constant
$\Lambda_0$, and the quadrupole anisotropy $Q_{rms}$ of the cosmic
microwave background radiation used for the
normalization of the linear spectrum. The parameters are quite
standard for these models. The linear transfer functions of the SCDM and
the LCDM models are taken from Bardeen et al. (1986) and that of the
CHDMs from Klypin et al. (1993). For all four models, we have assumed
the Harrison-Zel'dovich primordial power spectrum.
It is worth pointing out that the
$Q_{rms}$ normalization is compatible with the clustering of nearby galaxies
in the LCDM, CHDM1 and CHDM2 models, but is too high
for the SCDM model.

Part of the motivation for the two CHDM models comes
from the possible non-zero rest mass of
neutrinos. It has been claimed that the sum of the neutrino masses might be in
the range 3.5 eV $< \sum m_{\nu i}<$ 6 eV (e.g. Louis 1994). Therefore,
the allowed range for
the neutrino density would be $0.2\leq \Omega _{\nu} \leq 0.3$ when $h=0.5$.
The density parameters of the CHDM1 and CHDM2 models are taken to be
the maximum and minimum of this range.

\bigskip

\noindent{\it 3.1  Abundance of possible hosts of C IV systems}

\bigskip

The redshift evolution of the number density, $N_c(\geq M)$, of collapsed
halos with mass greater than a given $M$ are shown in
Figs. 4a-d. The eight curves in each figure correspond to
$M=10^{10+n0.5} \ M_{\odot}$, with $n=0,1... 7$ from top to bottom.

As discussed in \S 1, QSO metal absorption systems should be
hosted by collapsed halos, because only in such regions can
the formation of stars and evolution of heavy elements
be taking place. Therefore, the number of the observed C IV systems
can be used as a lower limit to the number of collapsed holes.

The number density of C IV systems is plotted as crosses in Fig. 4.
The 3-D densities in the middle of the diagram were deduced from the
observed line-of-sight number density (Steidel 1990) under
the assumption that each absorber is spherical and has a radius of
39 $h^{-1}$ kpc (Crotts et al. 1994). To consider the effect of uncertainties
in the radius,
we also plot in Fig. 4 the 3-D number densities if the radius is
5 times less than, and 5 times greater than 39 $h^{-1}$ kpc.
These are the upper set of points (8.0 $h^{-1}$ kpc) and the lower
set of points (195 $h^{-1}$ kpc), respectively. The factor of 5 might be
enough to account for uncertainties in the measurement of the radius.
It would be safely to use the data corresponding to the 195 $h^{-1}$kpc
radius as the lower limit to the real number.

 From Fig. 4, one sees that the possible hosts of C IV systems in the
SCDM and LCDM models should have a mass threshold of about
$10^{12.5}\ M_{\odot}$. Say,
that the number density of $M> 10^{12.5}\ M_{\odot}$
collapsed halos is great enough to fit with the observed C IV lines.
Because the lower the mass threshold, the more the collapsed halos, any mass
threshold less than $10^{12.5}\ M_{\odot}$ is allowed by the observation.
In the CHDM1 and CHDM2 models, the number densities of collapsed halos
are much less than those in the SCDM and LCDM models. In order to have
enough collapsed halos hosting C IV systems, one has to choose much smaller
mass thresholds. Figs. 4c and 4d showed that the assumed C IV halos
should have a mass threshold as low as
$10^{11}\ M_{\odot}$ if the 195 $h^{-1}$ kpc radius is used, or as
low as $10^{9.5}\ M_{\odot}$ for the 39 $h^{-1}$ kpc radius.
This mass threshold is about one and a
half orders of magnitude lower than that
in the SCDM and LCDM models. We conclude that in the CHDM models, the halos
that host C IV must have masses as, at least, low as $10^{11}\ M_{\odot}$.

\bigskip

\noindent{\it 3.2 Correlation functions of collapsed halos at
high redshifts}

\bigskip

Using the approximation of
Eq. (23), we present the correlation function
$\xi (r, \ge M)$ for collapsed halos in the SCDM, LCDM, CHDM1 and CHDM2
models in Fig. 5a-d. The mass threshold $M$ is taken to be $10^{10+n0.5}\
M_{\odot}$. The eight curves in each model correspond
to $n=0, ...7$ from left to right, respectively.

The histogram with error bars in Fig. 5 is the measured line-of-sight
correlation function of C IV absorption systems on scales from
$\Delta v \sim 300$ to 1,000 km s$^{-1}$ (Sargent et al. 1988).
C IV pairs with velocity differences less than 300 km s$^{-1}$
probably do not represent large-scale clustering, but the internal
structures of the host galaxies. The peculiar velocities of the C IV
absorbers may also significantly contaminate the spatial correlations
on scales less than the velocity dispersion. Therefore, the correlations
below 300 km s$^{-1}$ have not been shown in Fig. 5. Since the radii of the
halos (39 $h^{-1}$ kpc or even 195 $h^{-1}$ kpc), are much smaller than
the scales being studied ($\geq$ 300 km s$^{-1}$), the 3-D correlation
function is almost the same as that in 1-D. Therefore, one can directly
compare the observed 1-D correlation length with theory.

Fig. 5a shows that the correlation function of $M \geq 10^{12}M_{\odot}$
halos in the SCDM model provide a good fit to the observational data.
The halos are also well within the upper limit of the mass threshold,
$10^{12}M_{\odot}$, derived in the last section.
The same can be said for the LCDM model. In fact, SCDM and LCDM are
indistinguishable in the top-hat evolution, because, as we argued before,
a spherical mass has almost the same dynamic trajectory in all flat
universes with $\Lambda \leq 0.8$. Therefore,
with the proper parameters,
both SCDM and LCDM are consistent with the abundance and two-point
correlation function of C IV systems.

The two CHDMs are in trouble. According to the abundance fitting, the
hosting halos of C IV systems should have masses as small as $10^{11}
M_{\odot}$. However, the correlation function of such halos is much less
than is observed. In order to match the observed correlation function,
the halo masses must be at least $10^{12} M_{\odot}$ (Figs. 5c and
5d). But the number of these halos is too few to account for the
number of observed C IV systems. Therefore, there are no consistent
parameters for CHDM1 and CHDM2 under which both the abundance and
correlation tests can passed.

As pointed out by Heisler, Hogan \& White (1989), the observed 1-D
correlation function may not properly show the real-space correlation
but may be amplified somehow by the peculiar velocity. According to their
estimation based on a simple model, the amplitude of the correlation function
on scales less than the peculiar velocity could be amplified by the factor
of $(r_0 / r_{cl})^{\gamma}$, where $r_0$ is the correlation length,
$\gamma$ the correlation index and $r_{cl}$ the cloud radius. Namely, the
correlation length is amplified by the factor $(r_0 / r_{cl})$.
In the CHDM models, the correlation length of $10^{11} M_{\odot}$
halos is less than 0.1 Mpc at redshift 2.8, so the amplification factor
does not exceed 3. Therefore, this mechanism is far from resolving
the discrepancy which are found on scales of 3 - 5 h$^{-1}$ Mpc in the CHDMs.

\bigskip

\noindent{\bf 4. Discussion and conclusions}

\bigskip

Using the approximate expressions of correlation function and halo's
number density in the PS
formalism, we have derived constraints on the masses of collapsed
halos in four models of structure formation. Generally speaking, because
the number density of collapsed halos is inversely dependent on the
threshold mass above which the halos are selected, the abundance of C IV
systems sets up an upper limit to the threshold mass for their hosts. On
the other hand, the 2-point correlation function of collapsed halos is
positively dependent on the threshold mass, so the C IV clustering observation
sets up a lower limit to the threshold mass. Therefore, a model should
be considered inconsistent if the lower limit is found to be larger
than the upper limit. The two CHDM models are just such examples. Even
when the size of the C IV absorbers is taken to be 5 times larger than
that observed, the upper limit ($10^{11} M_{\odot}$)
provided by the abundance is still inconsistent with the lower limit
($10^{12} M_{\odot}$) inferred from the correlation.

We note from Fig. 3 that the mass correlation functions of CHDMs are much
less than those of SCDM and LCDM. Yet, their corresponding halo correlations
are greatly enhanced if we look at the same mass thresholds in Fig. 5.
This is due to the large gravitational bias of massive halos in the CHDM
models. Could it be possible that other biasing effects can further enhance
the correlations, so the models can finally pass the correlation test?
The answer probably is negative, because, expect gravitation, there are
almost no available bias mechanisms for generating a higher correlation
for less mass halos. Hydrodynamical processes generally are ineffective to
produce any inhomogeneity on scales equal to or larger than 5 h$^{-1}$Mpc, as
the streaming velocity of gaseous component is only about a thousand km
s$^{-1}$
on average.

Changing the density parameter of the hot dark matter from
$\Omega_{h}= 0.3$ (CHDM1)
to $0.2$ (CHDM2) does not relieve this difficulty much either. Though
CHDM2 has a variance 1.2 times that of CHDM1 in the linear evolution, it
still cannot produce enough halos hosting C IV systems.

Finally,
as pointed out in \S 2.4, our approximation of the correlation function may
give higher power on small scales than some N-body simulations; however,
this does not affect the conclusion about the CHDMs models. If Eq. (23)
really overestimates the correlation function on small scales, the
difference between the models and the observation would be even more
significant than is shown in Figs. 5c and 5d. In that case, the difficulty
for the models would become more severe.

\bigskip
\bigskip

{\em Acknowledgments}\ \ Many thanks are due to Houjun Mo for stimulating
discussions on the description and application of the Press-Schechter
formalism. We are grateful to Van Dixon for his help in improving the
quality of the manuscript. HGB is supported by a fellowship of the
World Laboratory.

\newpage

\noindent{\bf References}

\bigskip

\noindent$^*$ World Lab Fellowship, Beijing Astronomical Observatory,
Beijing, P.R. China

\ref Bardeen, J.M., Bond, J.R., Kaiser, N. \& Szaley, A.S. 1986, ApJ,
    304, 15

\ref Bechtold, J., Crotts, P.S., Duncan, R.C. \& Fang Y. 1994, ApJ, 437, L83

\ref Bergeron, J. et al. 1994, ApJ, 436, 33

\ref Bi, H.G., Ge, J. \& Fang, L.Z. 1995, ApJ, 452, 90

\ref Bond, J.R., Cole, S., Efstathiou, G. \& Katser, N. 1991, ApJ, 379, 440

\ref Crotts, A.P.S., Bechtold, J., Fang, Y. \& Duncan, R.C. 1994, ApJ, 437, L79

\ref Dinshaw, N., Impey, C.D., Foltz, C.B., Weymann, R.J. \& Chaffee, F.M.,
1994, ApJ, 437, L87

\ref Efstathiou, G., Frenk, C.S., White, S.D.M. \& Davis, M. 1988,
     MNRAS, 235, 715

\ref Heisler, J., Hogan, C.J. \& White, S.D.M. 1989, ApJ 347, 52

\ref Hamilton, A.J.S., Kumar, P., Lu, E. \& Matthews, A. 1991, ApJ, 374, L1

\ref Jing, Y.P. \& Fang, L.Z. 1994, ApJ, 432, 438

\ref Kaiser, N. 1984, ApJ, 284, L9

\ref Klypin, A., Holtzman, J., Primack,J. R. \& Regos, E. 1993, ApJ, 416, 1

\ref Klypin, A., Borgani, S., Holtzman, J. \& Primack, J. 1995, ApJ, 444, 1

\ref Lacey, C., \& Cole, S., 1994, MNRAS, 271, 676

\ref Louis W.C., 1994, in Proceedings of the XVI Conference on Neutrino
physics and Astrophysics, Eilat, Israel.

\ref Ma, C.P. \& Bertschinger, E. 1994, ApJ, 434, L5

\ref Mo, H.J. \& Miralda-Escud\'{e}, J. 1994, ApJ, 430, L25

\ref Mo, H.J. \&  White, S. 1995, preprint

\ref Peebles, P.J.E. 1984, ApJ, 284, 439

\ref Press, W.H. \& Schechter, P. 1974, ApJ, 187, 425

\ref Sargent, W.L., Boksenberg, A. \& Steidel, C.C. 1988, ApJS, 68, 539

\ref Steidel, C.C. 1990, ApJS, 72, 1

\ref Wolfe, A. 1993, in {\it Relativistic Astrophysics and Particle
   Cosmology}, eds. Akerlof, C.W. \& Srednicki, M.A., N.Y. Academy of
Sciences

\newpage

\begin{center}

{\bf Table 1. Parameters of the cosmological models}

\bigskip
\bigskip

\begin{tabular}{l|l|l|l|l}
\cline{1-5}
             &       &      &       &       \\
             &  SCDM & LCDM & CHDM1 & CHDM2 \\
             &       &      &       &       \\
\cline{1-5}
             &       &      &       &       \\
$h$          &  0.5  & 0.75 & 0.5   & 0.5   \\
$\Omega_{cb} $ &  1    & 0.3  & 0.7   & 0.8   \\
$\Omega_h$ &  0    & 0    & 0.3   & 0.2   \\
$\Lambda_0$  &  0    & 0.7  & 0     & 0     \\
$Q_{rms}(\mu K)$   &  18.0 & 18.0    & 18.0  & 18.0  \\
             &       &      &       &       \\
\cline{1-5}
\end{tabular}

\end{center}

\newpage

\noindent{\bf Figure Captions}

\begin{description}

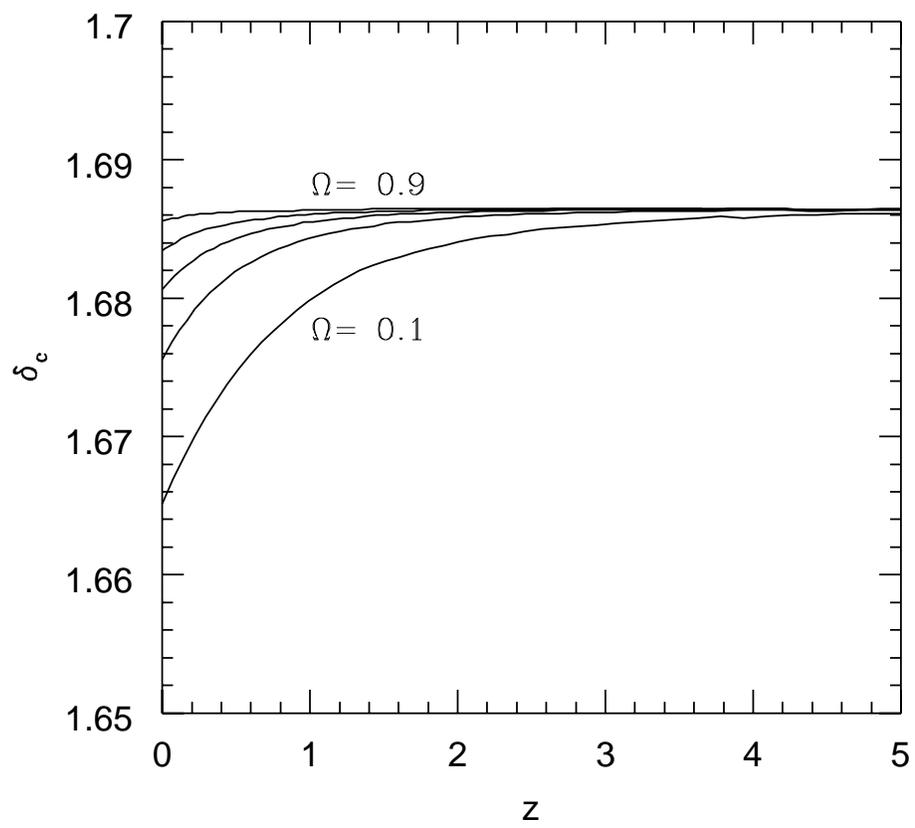
\item[{\bf Figure 1}] \ Threshold $\delta_c$ as a function of the collapsing
redshift $z_c$ for $\Omega =$ 0.9, 0.7, 0.5, 0.3, 0.1, plotted from
top to bottom.

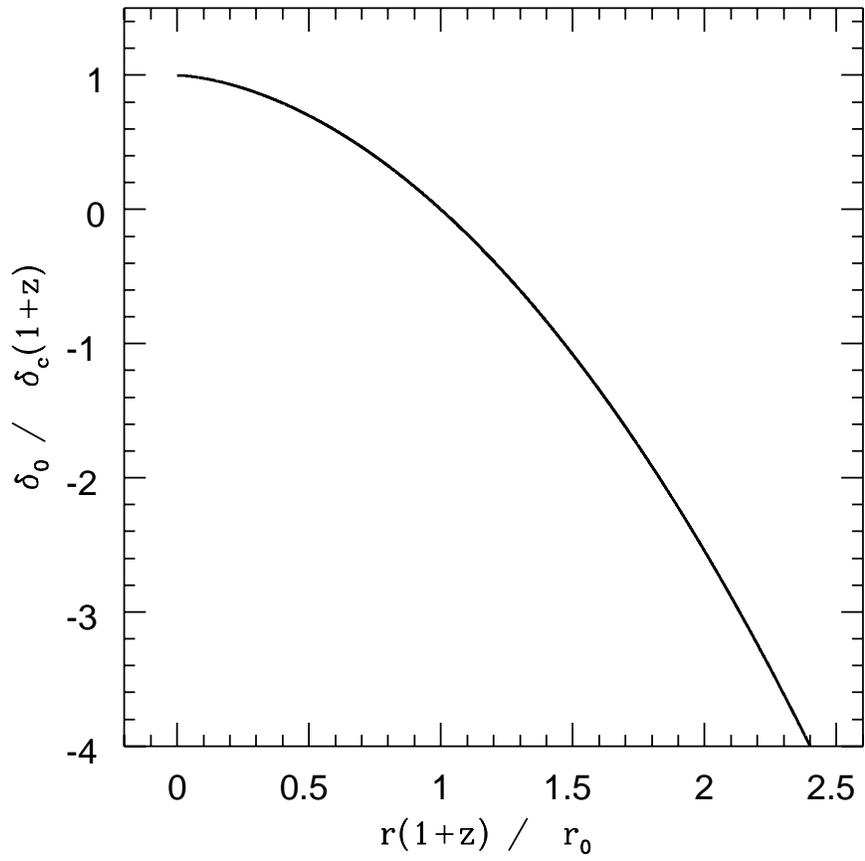
\item[{\bf Figure 2}] \ The function $f(x)$ defined by $\delta _0 = \delta
_c (1+z) f ( r(1+z)/r_0 )$ according to Eqs. (14) and (15).

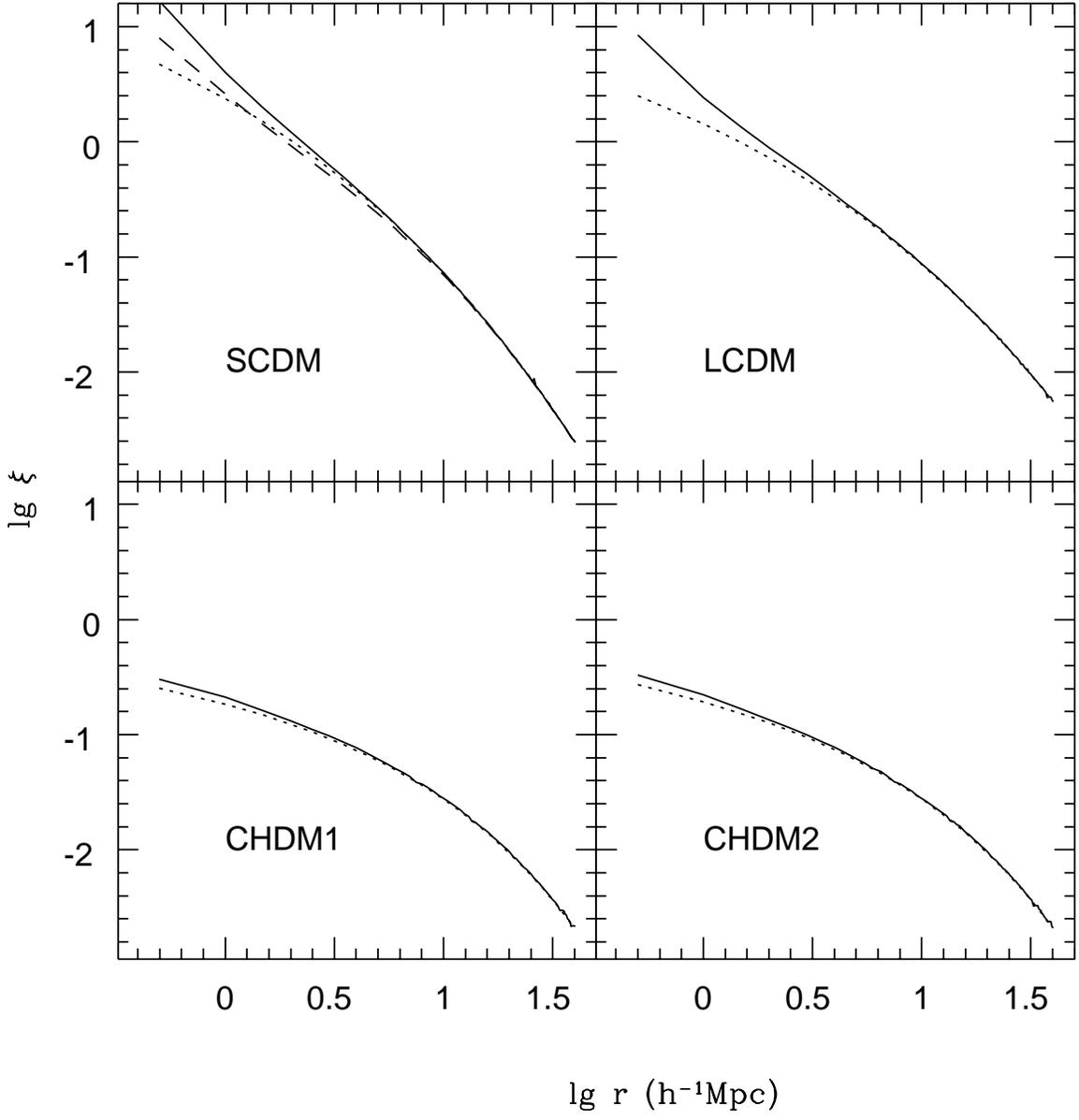
\item[{\bf Figure 3}] \ The integrated mass correlation functions in the
SCDM, LCDM, CHDM1 and CHDM2 models.  The solid lines are calculated from
Eqs.(19) and (20). The dotted lines are the linear correlation functions,
and the dashed line in the SCDM model is from the empirical formula of
Hamilton et al. (1991).

\item[{\bf Figure 4a}] \  The comoving abundance of collapsed halos with
masses greater than $M$ in the SCDM model. The eight curves correspond to
$M = 10^{10 + 0.5n} M_{\odot}, \ \  n=0,1,...,7$, from top to
bottom. The bold
error bars in the middle are the number densities of QSO C IV absorption
systems deduced from 1-D observational data and the measured radius 39
$h^{-1}$ kpc. The data sets above and below the middle ones are the number
densities when the radius is ($39/5)$ $h^{-1}$ kpc and ($5 \times 39)$
$h^{-1}$ kpc, respectively.

\item[{\bf Figure 4b}] \ The same as Fig. 4a for the LCDM model.

\item[{\bf Figure 4c}] \ The same as Fig. 4a for the CHDM1 model.

\item[{\bf Figure 4d}] \ The same as Fig. 4a for the CHDM2 model.

\item[{\bf Figure 5a}] \ The correlation function of collapsed halos
with masses greater than $M$ in the SCDM model. The eight curves
correspond to $M=10^{10 + 0.5n}$ M$_{\odot}$, $n=0,1,...,7$,
from left to right. The histogram with error bars
is the measured correlation
function of C IV systems.

\item[{\bf Figure 5b}] \ The same as Fig. 5a for the LCDM model.

\item[{\bf Figure 5c}] \ The same as Fig. 5a for the CHDM1 model.

\item[{\bf Figure 5d}] \ The same as Fig. 5a for the CHDM2 model.

\end{description}

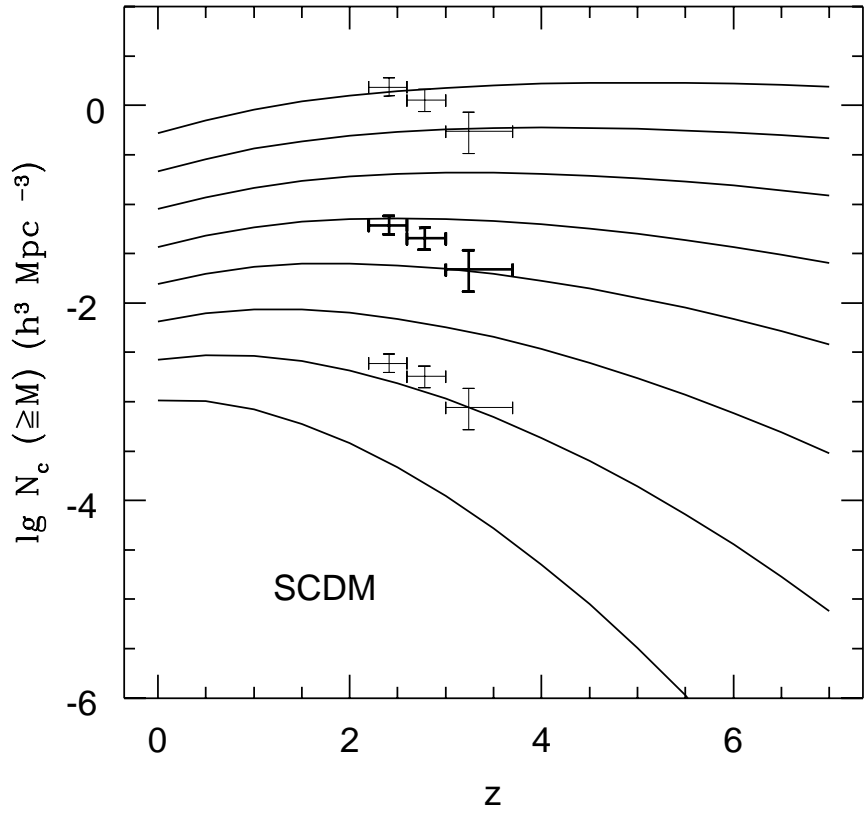
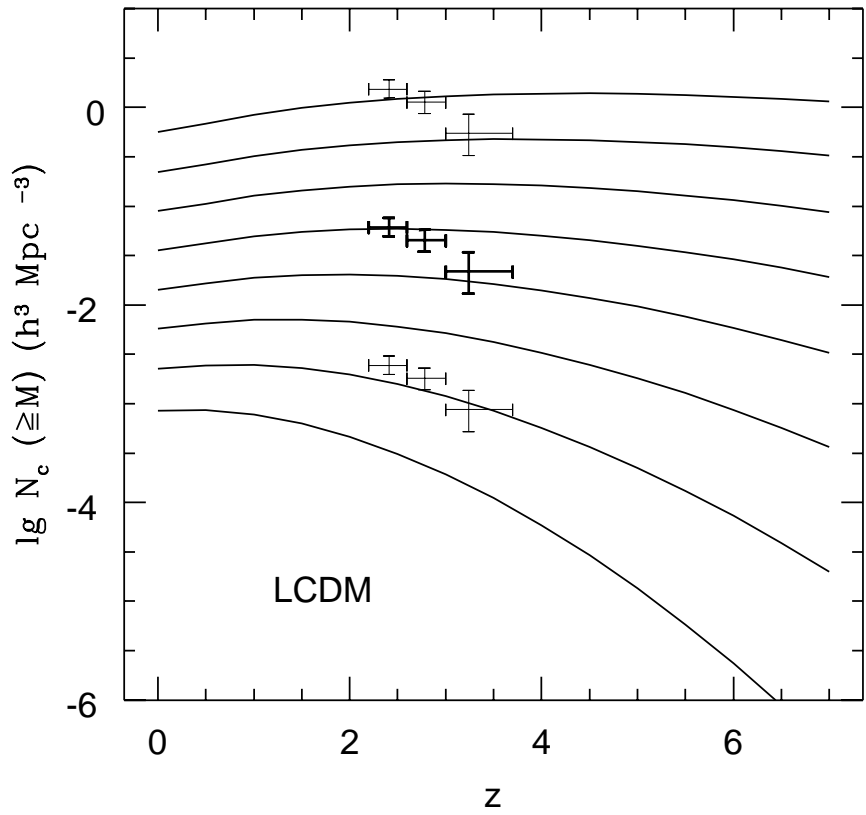
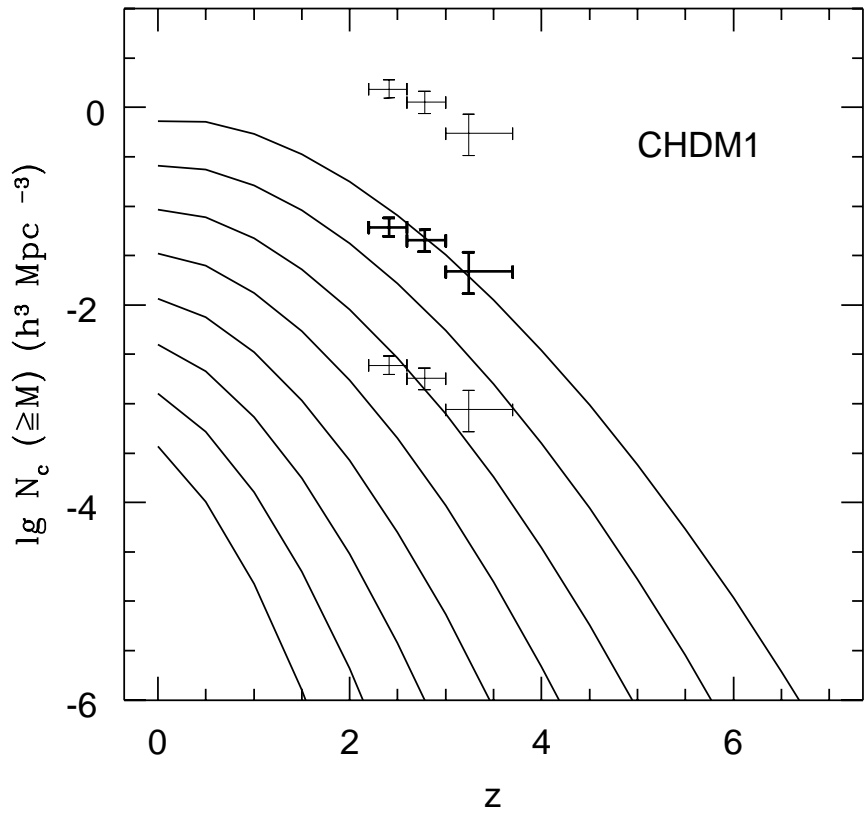
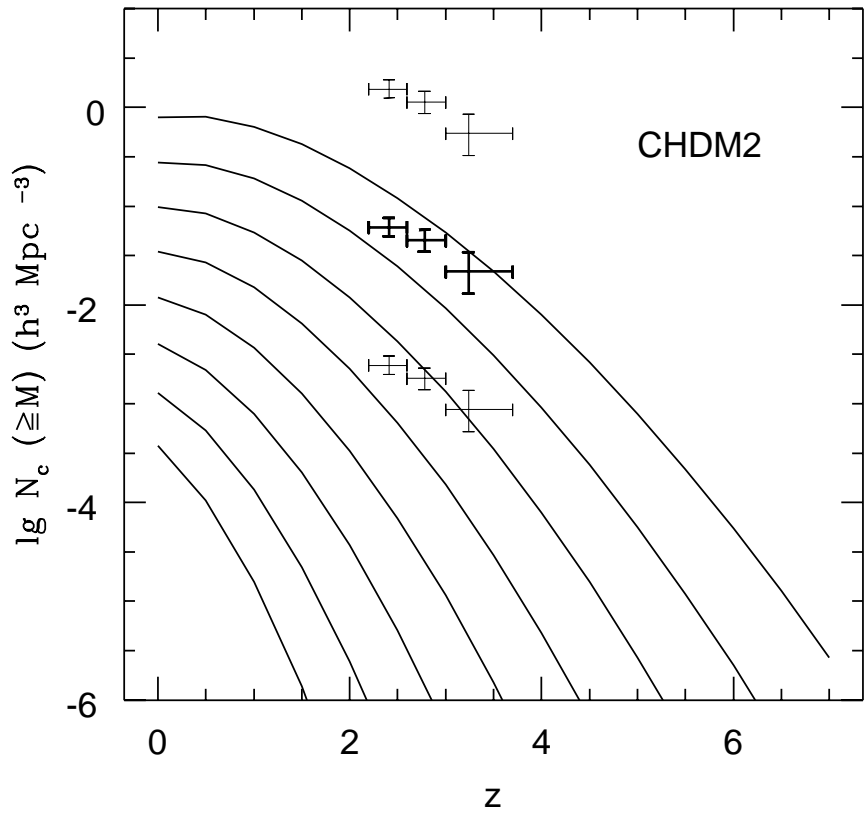
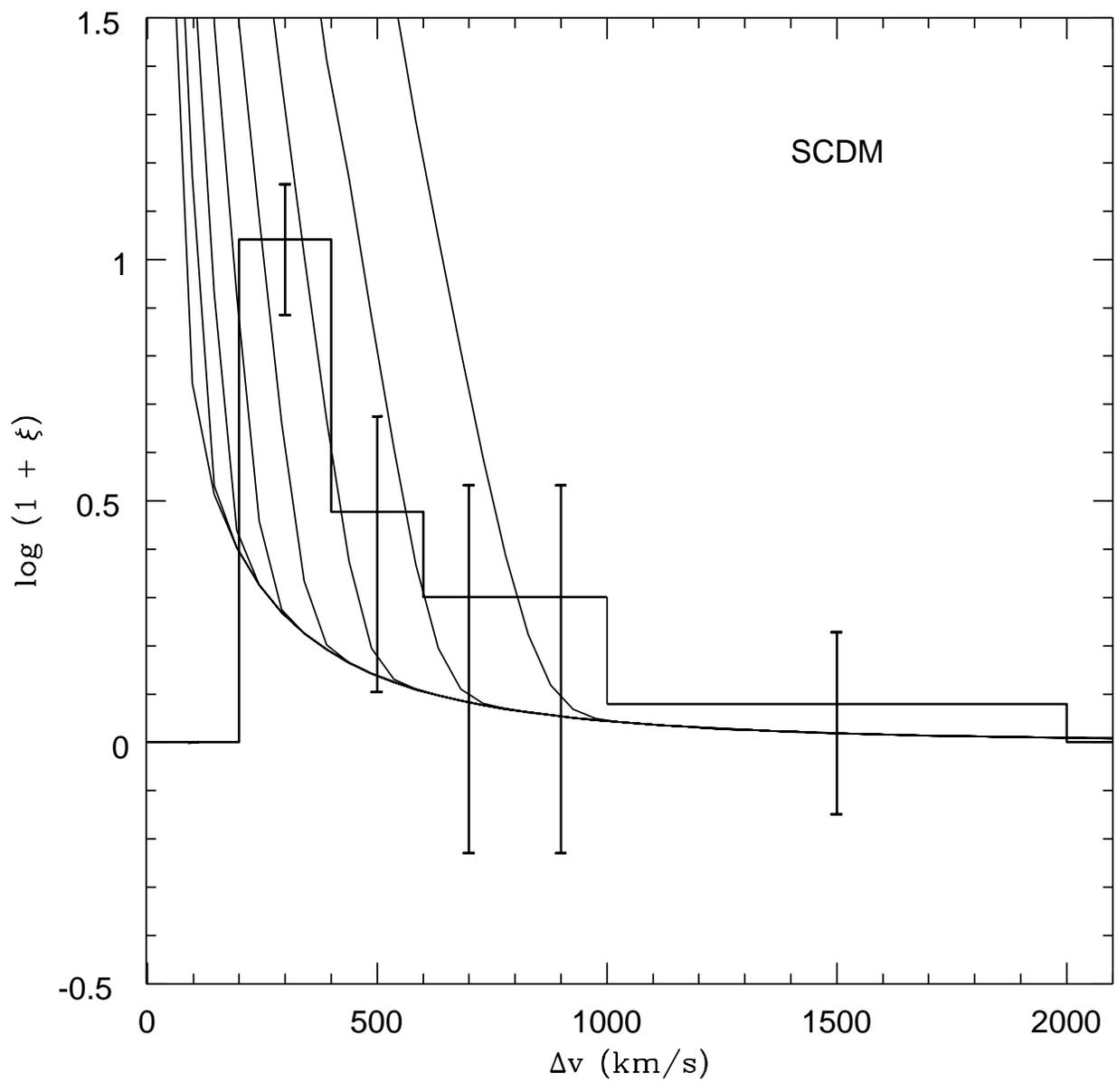
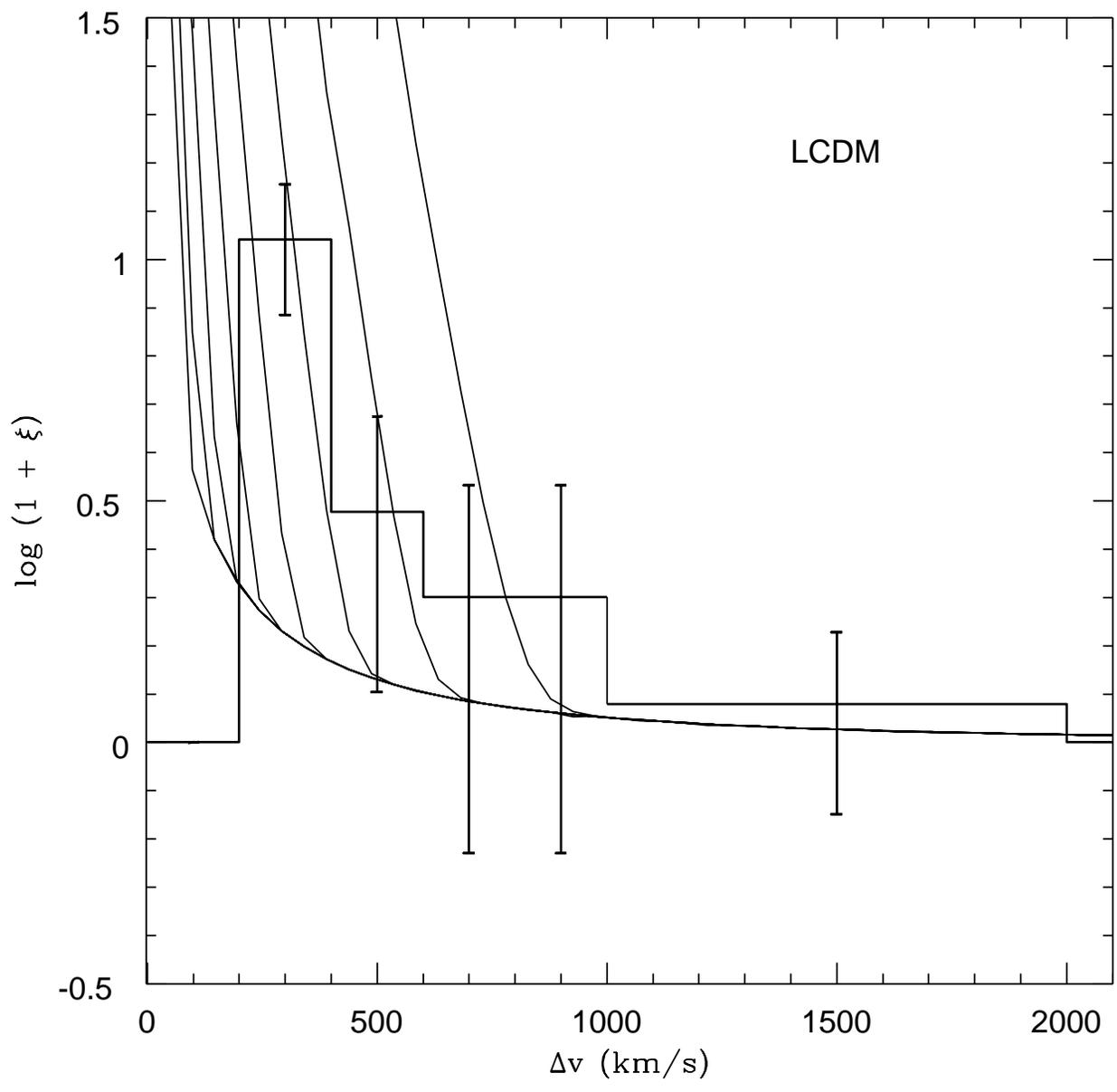
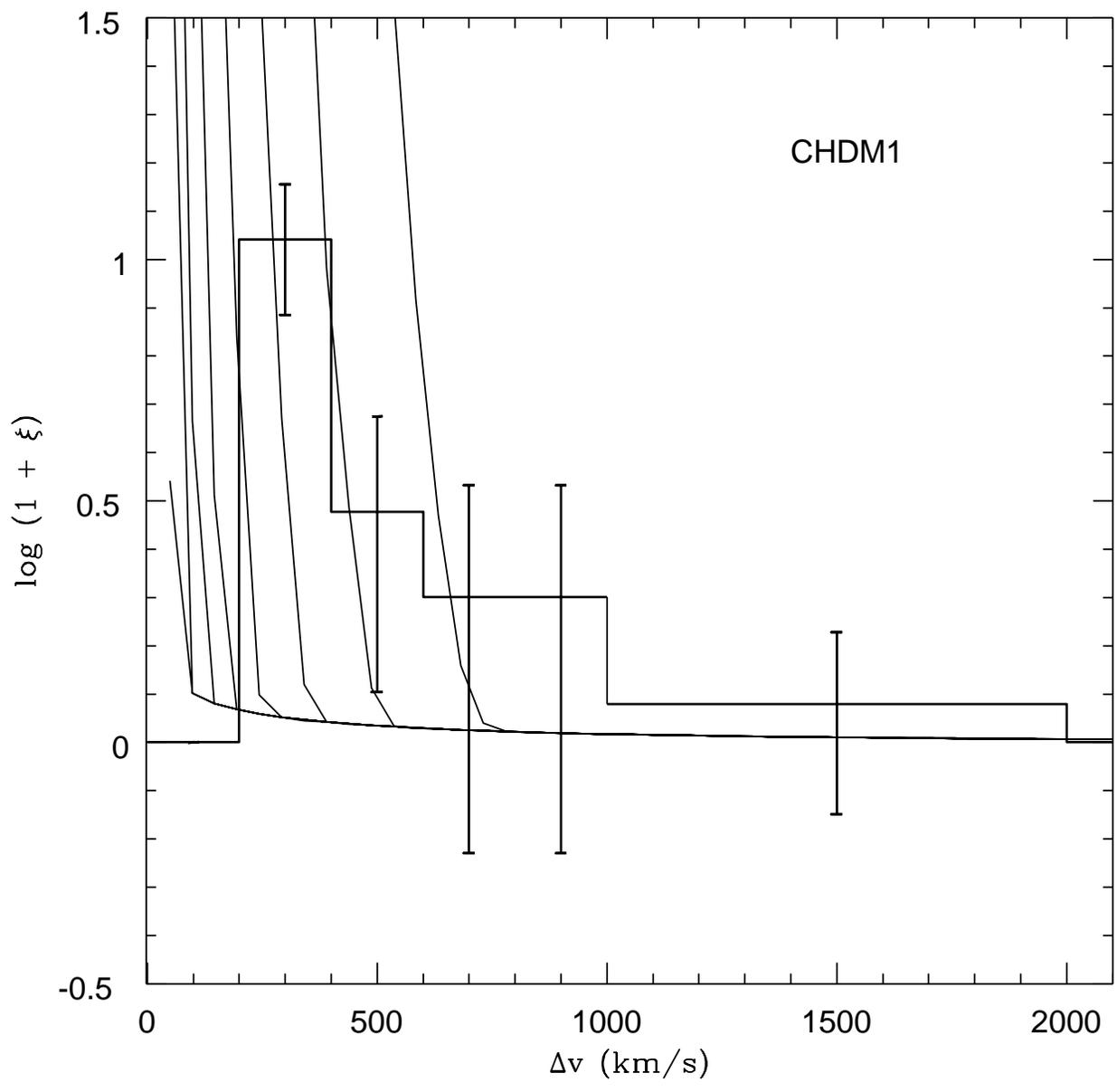
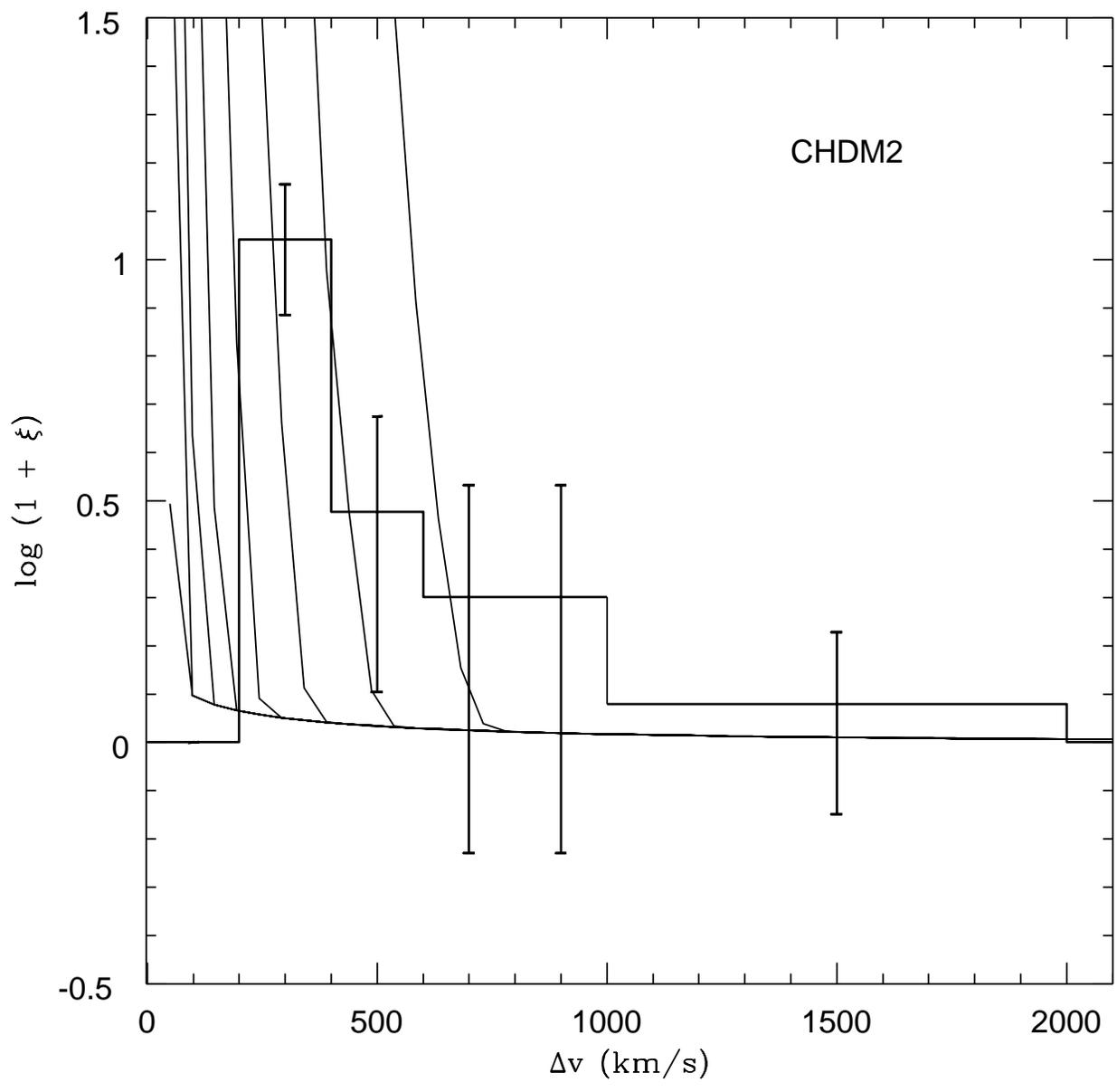
\end{document}